\documentclass{moriond}

\usepackage{amsmath}

\def\Journal#1#2#3#4{{#1} {\bf #2}, #3 (#4)}

\def\AeA{\em Astronomy and Astrophysics}
\def\ASR{\em Advances in Space Research}
\def\LRR{\em Living Review in Relativity}
\def\MNRAS{\em MNRAS}
\def\NAT{\em Nature}

\def\PRD{{\em Phys. Rev.} D}

\def\be{\begin{equation}}
\def\ee{\end{equation}}
\def\bea{\begin{eqnarray}}
\def\eea{\end{eqnarray}}

\newcommand{\Photo}{}

\begin{document}
\vspace*{4cm}
\title{GRAVITATIONAL WAVES RADIATED BY MAGNETIC GALACTIC BINARIES AND DETECTION BY LISA}

\author{A. BOURGOIN${}^{1,2}$, C. LE PONCIN-LAFITTE${}^{1}$, S. MATHIS${}^{2}$, and M.-C. ANGONIN${}^{2}$}

\address{${}^{1}$SYRTE, Observatoire de Paris, PSL Research University, CNRS, Sorbonne Universités, UPMC Univ. Paris 06, LNE, 61 avenue de l'Observatoire, 75\,014 Paris, France\\${}^{2}$Département d'Astrophysique-AIM, CEA/IRFU/DAp, CNRS/INSU, Université Paris-Saclay, Universités de Paris, Orme des Merisiers, 91\,191 Gif-sur-Yvette, France}

\maketitle\abstracts{
In the context of the future Laser Interferometer Space Antenna (LISA) mission, galactic binary systems of white dwarfs and neutron stars will represent the dominant source of Gravitational Waves (GWs) within the $10^{-4}-10^{-1}\ \mathrm{Hz}$ frequency band. It is expected that LISA will measure simultaneously, the GWs from more than ten thousands of these compact galactic binaries. The analysis of such a superposition of signals will represent one of the greatest challenge for the mission. Currently, in the LISA Datacode Challenge, each galactic binary is modeled as a quasi-monochromatic source of GWs. This corresponds to the circular motion of two point-masses at the 2.5 post-Newtonian approximation. If this picture is expected to be an accurate description for most of the galactic binaries that LSIA will detect, we nevertheless expect to observe eccentric systems with complex physical properties beyond the point-mass approximation. In this work, we investigate how a binary system of highly magnetic objects in quasi-circular orbit could affect the quasi-monochromatic picture of the GW signal detected by LISA. We demonstrate that the eccentricity generates additional frequency peaks at harmonics of the mean motion and that magnetism is responsible for shifting each frequency peak with respect to the case without magnetism. We provide analytical estimates and argue that LISA will be able to detect magnetism if it can measure the main peaks at two and three times the mean motion with a sufficient accuracy.}

\section{Introduction}

Laser Interferometer Space Antenna (LISA) is the ESA L-class mission dedicated to the observation of Gravitational Waves (GWs) from space \cite{2017arXiv170200786A} in the frequency band from below $10^{-4}\ \mathrm{Hz}$ to above $10^{-1}\ \mathrm{Hz}$. Within this frequency range the main source of GWs are the Galactic Binaries (GBs) which comprise White Dwarfs (WDs) and Neutron Stars (NSs) in various combination. In LISA's bandwidth the typical orbital period of GBs is ranging from minutes to several hours which corresponds to semi-major axis between $10^4\ \mathrm{km}$ to $10^6\ \mathrm{km}$. Therefore, LISA will observe GBs during the inspiral phase, before the merger which can be detected by ground-based instruments. Around ten thousands of inspiral GBs should be resolved by LISA~\cite{PhysRevD.73.122001}. Among these, tens of them are of a particular interest since they are guaranteed sources for LISA, with an expected time scale of detection of the order of few weeks. These are called the ``verification binaries'' and are already identified as GW sources with a high signal-to-noise ratio within LISA's bandwidth. They will serve for calibrating the detector, thus any mis-modeling while processing the GW signal from verification binaries can potentially have an impact on the other extra-galactic scientific objectives of LISA. Currently within the LISA Datacode Challenge (LDC) \cite{LDCGroup}, GBs are modeled as quasi-monochromatic sources of GWs. This corresponds to the circular motion of two point-masses at the 2.5 Post-Newtonian (PN) approximation \cite{2014LRR....17....2B}. However, it is expected that eccentric GBs will be detected within the frequency band of the LISA mission \cite{2020MNRAS.491.3000M}. Furthermore, we could also expect to observe a more complex signal than the monochromatic approximation as a consequence of perturbations due, for instance, to internal processes occurring inside WDs and NSs. As a matter of fact, WDs and NSs are among the most magnetized objects in the universe, with magnetic fields that can reach up to $10^9\ \mathrm{G}$ for WDs and up to $10^{15}\ \mathrm{G}$ for NSs \cite{2020AdSpR..66.1025F}.

\section{Origin of magnetism in white dwarfs and neutron stars}

It is expected that $20\%$ of the total WD population \cite{2020AdSpR..66.1025F} possess strong magnetic fields in the range between $10^6\ \mathrm{G}-10^9\ \mathrm{G}$, while highly magnetic NSs with magnetic fields between $10^{14}\ \mathrm{G}-10^{15}\ \mathrm{G}$ (i.e., the magnetars) should represent around $10\%$ of the total NS population \cite{2008MNRAS.387..897T}. The origin of these strong magnetic fields is still debated and many scenarios have been proposed but none of them can fully explain the all set of electromagnetic (EM) observations so far \cite{2021MNRAS.507.5902B}.

The ``merging scenario'' states that highly magnetic WDs or NSs are formed from the merger of a binary system consisting of a preliminary WD and a mass transferring companion. The main observational motivation comes from the fact that highly magnetic WDs and magnetars are mainly observed either isolated or in cataclysmic variable but not in binary systems with a detached low-mass main-sequence companion \cite{2008MNRAS.387..897T}. Nevertheless, highly magnetic objects in binary system with a detached companion may be rare but do exist as pointed out by Bagnulo \emph{et al} \cite{2020A&A...634L..10L}.


The ``fossil-fields'' scenario states that the emergence of highly magnetic fields results from a mechanism of magnetic flux conservation during the stellar evolution off the main-sequence~\cite{2005MNRAS.356..615F}. The main observational motivation in favor of the ``fossil-field'' hypothesis comes from the range of magnetic fields in Ap/Bp stars, and in some stars of spectral type O, that matches perfectly the range of magnetic fields observed in highly magnetic WDs and NSs, respectively~\cite{2005MNRAS.356..615F,2005MNRAS.356.1576W}. This scenario is an attractive possibility, however, it cannot properly explain why highly magnetic objects are mostly observed individually and not paired with a detached companion. Nevertheless, numerical investigations by Braithwaite {\it et al} \cite{2004Natur.431..819B} seem to favor the ``fossil-fields'' scenario. Indeed, magnetohydrodynamic simulations have shown that a stable dipolar magnetic field can develop from an arbitrary initial magnetic field and persist over the lifetime of the compact stars without requiring a preliminary merger.

By complementing the EM observations, LISA will potentially bring precious informations about the nature of magnetism inside WDs and NSs. To do so, the influence of the magnetism on the GW signal must be properly investigated for the lifetime of the LISA mission.

\section{Secular dynamics of magnetic compact binaries}

\begin{figure}
\begin{minipage}{0.5\linewidth}
\centerline{\includegraphics[width=0.98\linewidth]{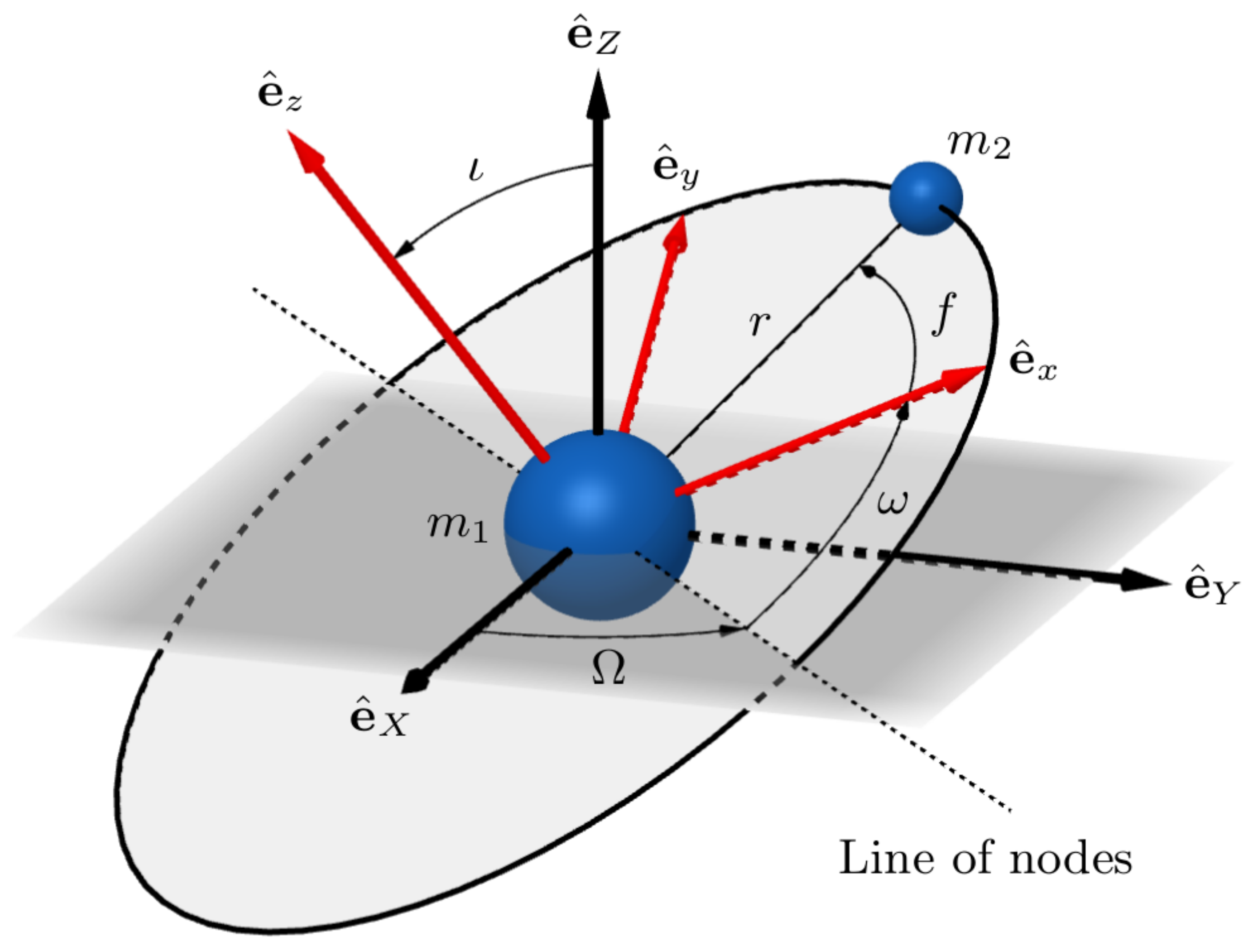}}
\end{minipage}
\hfill
\begin{minipage}{0.5\linewidth}
\centerline{\includegraphics[width=0.98\linewidth]{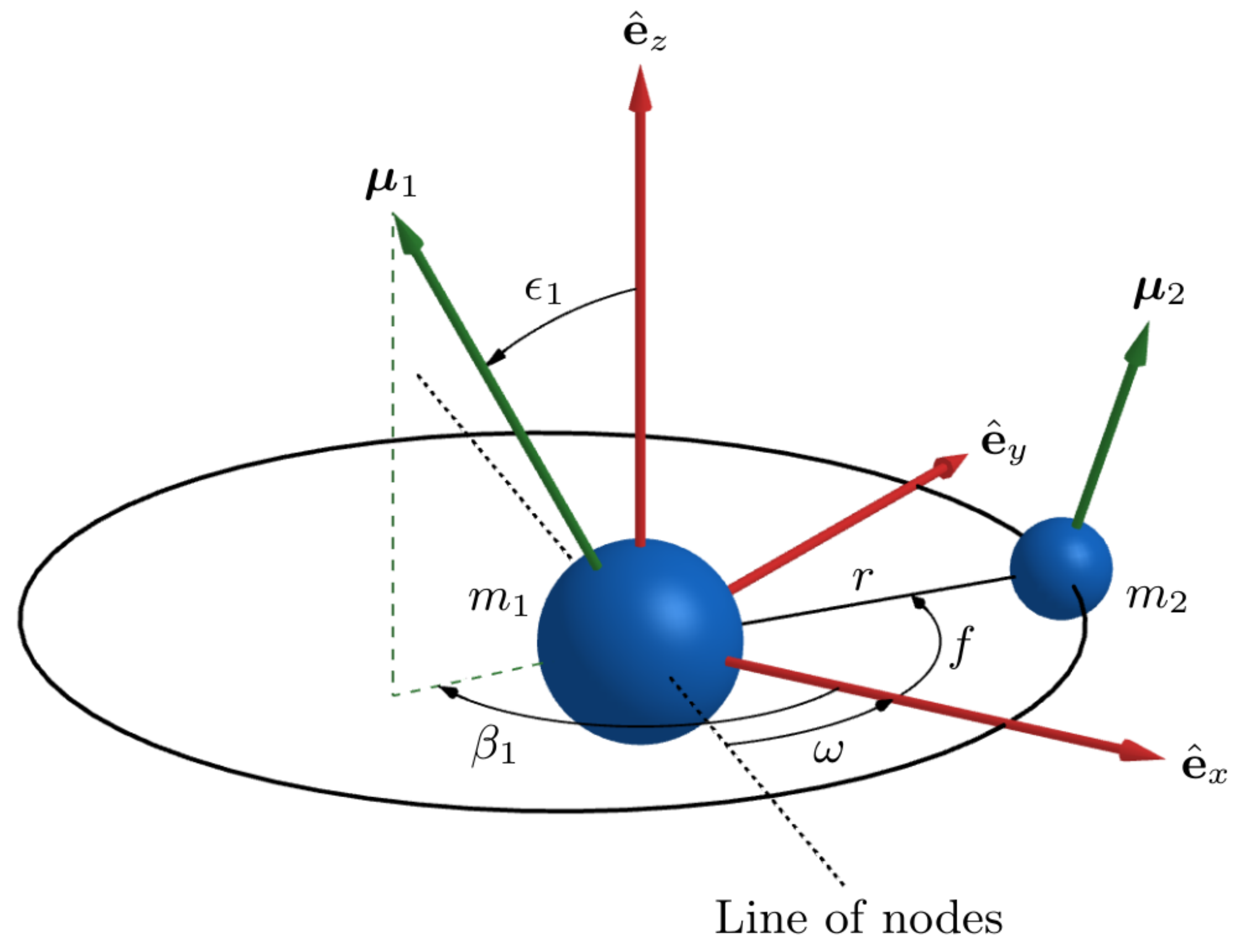}}
\end{minipage}
\caption[]{Orientation of the orbit (left). Orientation of the magnetic moments (right).}
\label{fig:orbit}
\end{figure}

In a recent work \cite{arxiv2201.03226}, we have derived analytically first-order analytical estimates for the secular evolution of the rotational and orbital motion of a compact binary system of total mass $m=m_1+m_2$, where $m_1$ and $m_2$ are the masses of the primary and secondary, respectively. We considered a relativistic description of the motion up to the 2.5PN approximation in order to be coherent with the loss of orbital energy and angular momentum that is carried away from the source by GWs. We considered magnetism of both stars through the dipole-dipole interaction. In agreement with the ``fossil-fields hypothesis'', we assumed static dipolar magnetic fields for both stars. In addition, while focusing on the inspiral phase, we also assumed that the internal electric currents that generate the magnetic fields are not significantly distorted by the magnetic field of the companion. We considered, for simplicity, that the directions of the magnetic moments of both stars (called $\mu_1$ for the primary and $\mu_2$ for the secondary) are aligned with their spin's axis. Within this framework, we derived the secular equations of motion for the evolution of orientations of the magnetic moments and for the orbit of the binary system. We used the Lagrange planetary equations for non-singular orbital elements, namely $(a,z,\zeta,L)$, where $a$ is the semi-major axis, $z$ is the imaginary eccentricity vector ($z=e\,\mathrm{e}^{\mathrm{i}\varpi}$ with $e$ the eccentricity and $\varpi$ the longitude of the pericenter, i.e., $\varpi=\Omega+\omega$, where $\Omega$ is the longitude of the ascending node and $\omega$ is the argument of the pericenter), $\zeta$ is the imaginary inclination vector ($\zeta=\sin(\iota/2)\,\mathrm{e}^{\mathrm{i}\Omega}$ with $\iota$ the inclination), and $L$ is the mean longitude ($L=\varpi+M$ with $M$ the mean anomaly) (see figure \ref{fig:orbit}). The so-obtained equations are valid even for small eccentricity and small inclination. First-order solutions for the orbital motion $(a(t),z(t),\zeta(t),L(t))$ are as follows:
\begin{subequations}
\begin{align}
  a(t)=a_0&+\dot{a}_{\mathrm{GR}}t\text{,}\qquad z(t)=e_0\,\mathrm{e}^{\dot e_{\mathrm{GR}}t/e_0}\mathrm{e}^{\mathrm i\varpi(t)}\text{,}\qquad\zeta(t)=\zeta_0+\widetilde{\zeta}_{\mathrm{M}}(t)\text{,}\\
  L(t)&=L_0+\widetilde{L}_{\mathrm{M}}(t)+(n_0+\dot L_{\mathrm{GR}}+\dot L_{\mathrm{M}})t-\frac{3n_0}{4a_0}\,\dot a_{\mathrm{GR}}t^2\text{,}
\end{align}
\end{subequations}
where the longitude of the pericenter is $\varpi(t)=\varpi_0+\widetilde{\varpi}_{\mathrm{M}}(t)+(\dot\varpi_{\mathrm{GR}}+\dot \varpi_{\mathrm{M}})t$. The ``tilde'' denotes a periodic contribution while a ``dot'' is used for a secular rate of change. The subscripts ``M'' and ``GR'' are employed to denote a magnetic effect and a relativistic contribution (not shown here), respectively. The subscript ``$0$'' is employed to denote an initial condition (e.g., $a_0=a(t=0)$). The parameter $n_0$ is the initial mean motion. The main contribution to retain from the magnetic interaction is the secular drift on the mean longitude which scales as
\begin{equation}
  \dot L_{\mathrm{M}}=\frac{3\mu_0}{4\pi\sqrt{G}}\,\frac{\sqrt{m}}{m_1m_2}\,\frac{\mu_1\mu_2}{a_0{}^{7/2}}\,\frac{\left(1+\sqrt{1-e_0{}^2}\right)}{(1-e_0{}^2)^2}\cos\epsilon_{10}\cos\epsilon_{20}\text{,}
  \label{eq:dLM}
\end{equation}
where $\epsilon_{10}$ and $\epsilon_{20}$ are the obliquities of the magnetic moments of primary and secondary, respectively. $G$ is the gravitational constant and $\mu_0$ is the permeability of vacuum. In addition, we mention the secular contribution on the longitude of the pericenter which is used hereafter and which reads as follows: $\dot\varpi_{\mathrm{M}}=\dot L_{\mathrm{M}}/(1+\sqrt{1-e_0{}^2})$.

Equation \ref{eq:dLM} shows that a system of double WD in compact orbit is more likely to feel the effect of the magnetic interaction since it is proportional to $\mu_1\mu_2$  and evolves as the inverse of the semi-major axis to the power 7/2. (Magnetic moments are higher for WDs than for NSs \cite{arxiv2201.03226}).

\section{Gravitational mode polarizations}

For a binary system of two point-masses, the Einstein's quadrupole formula can be expressed in term of the regular orbital elements. The mode polarizations $h_+$ and $h_\times$ are conveniently given by the following Fourier's decomposition:
\begin{equation}
  h_+(t)-\mathrm{i}h_\times(t)=h(a(t))\sum_{k=-\infty}^{+\infty}c_k(z(t),\zeta(t))\,\mathrm{e}^{\mathrm{i}kL(t)}\text{,}
  \label{eq:modpol}
\end{equation}
where $c_k$ are the Fourier's coefficients ($|k|=2$ terms are $\propto e^0$, $|k|=1,3$ terms are $\propto e^1$, $|k|=2,4$ terms are $\propto e^2$, $\ldots$). The amplitude is $h(a)=2\eta(a/D)(Gm/c^2a)^2$ with $\eta=m_1m_2/m^2$ the mass parameter and $D$ the distance between the source and the field points.

After inserting the orbital first-order secular solutions (derived in the previous section) into Eq. \eqref{eq:modpol}, we derive the expression for the mode polarizations at zeroth-order in eccentricity (we provide the explicit form of the ``$+$'' polarization): $h_+{}^{\!\!(0)}(t)=h_0(1+\cos\iota_0)\cos (\phi^{(0)}+\Phi^{(0)}t+\dot\Phi^{(0)}t^2)$, where $\phi^{(0)}=2L_0$, and where the frequency and the frequency shift are given by
\begin{equation}
  \Phi^{(0)}=2n_0\bigg(1+\frac{\dot L_{\mathrm{GR}}}{n_0}+\frac{\dot L_{\mathrm{M}}}{n_0}\bigg)\text{,}\qquad\dot\Phi^{(0)}=-\frac{3n_0}{2a_0}\,\dot a_{\mathrm{GR}}\text{.}
\end{equation}

Thus, the form of the mode polarization $h_+{}^{\!\!(0)}$ is actually similar to LDC's. Therefore, at zeroth-order in eccentricity the LDC modeling already accounts for magnetic binary cases. However, if one wants to interpret the measured frequency $\Phi^{(0)}$ in term of the physical content (e.g., the masses), one has to consider magnetic corrections if $\sigma_{\Phi^{(0)}}<\Phi^{(0)}\dot L_{\mathrm{M}}/n_0$, where $\sigma_{\Phi^{(0)}}$ is the uncertainty of the measured frequency. Considering a binary of WDs with magnetic fields at the level of $B_1=B_2=10^{9}\ \mathrm{G}$, we find $\sigma_{\Phi^{(0)}}<7\times 10^{-8}\ \mathrm{Hz}$ for $\Phi^{(0)}=10^{-1}\ \mathrm{Hz}$. Some of the verification binaries are already known with a precision at the level of $10^{-10}\ \mathrm{Hz}$.

At first-order in eccentricity, the magnetic perturbation translates into the following additional signal (we provide the explicit form of the ``$+$'' polarization): $h_+{}^{\!\!(1)}(t)=(9e_0/4)h_0(1+\cos\iota_0)\cos(\phi^{(1)}+\Phi^{(1)} t+\dot\Phi^{(1)}t^2)+\ldots$ The phase reads as $\phi^{(1)}=3L_0-\varpi_0$, and the frequency and the frequency shift of this new component in the GWs signal are given by
\begin{equation}
  \Phi^{(1)}=3n_0\bigg(1+\frac{3\dot L_{\mathrm{GR}}-\dot\varpi_{\mathrm{GR}}}{3n_0}+\frac{3\dot L_{\mathrm{M}}-\dot\varpi_{\mathrm{M}}}{3n_0}\bigg)\text{,}\qquad\dot\Phi^{(1)}=-\frac{9n_0}{4a_0}\,\dot a_{\mathrm{GR}}\text{.}
\end{equation}
The ``ellipses'' in the expression of $h_+{}^{\!\!(1)}$ represent first-order terms in eccentricity that have a frequency different than $\Phi^{(1)}$. These terms have a smaller amplitude than $\Phi^{(1)}$. 

We see that the form of the mode polarization $h_+{}^{\!\!(1)}$ is actually similar to the LDC's zeroth-order term in eccentricity. It will thus be interpreted as a different source of GWs than the zeroth-order component, whereas it is just an harmonic at $\sim 3n_0$. If this secondary frequency peak, $\Phi^{(1)}$, can be detected and measured by LISA, the following linear combination: $1.5\Phi^{(0)}-\Phi^{(1)}=\dot\varpi_{\mathrm{GR}}+\dot\varpi_{\mathrm{M}}$, should permit to fully extract the magnetic fields information since $\dot\varpi_{\mathrm{M}}\propto \mu_1\mu_2/a_0{}^{7/2}$. Let us emphasize that the contribution from general relativity is well known and can be easily modeled at 1PN order: $\dot\varpi_{\mathrm{GR}}=3n_0(Gm/c^2p_0)$ with $p_0=a_0(1-e_0{}^2)$.

\vspace{-0.1cm}

\paragraph*{Acknowledgments.} This work was supported by the Programme National GRAM of CNRS/INSU with INP and IN2P3 co-funded by CNES and by CNES LISA grants at CEA/IRFU.

\end{document}